\documentclass[sigconf]{acmart}

\usepackage{times}

\usepackage{hyperref}
\usepackage{url}
\usepackage{xcolor}

\long\def\ignore#1{}

\title{Yes, BM25 is a Strong Baseline for Legal Case Retrieval}

\setcopyright{rightsretained}
\acmConference[COLIEE 2021]{COLIEE 2021 workshop: Competition on Legal
   Information Extraction/Entailment }{June 21, 2021}{Online}
\acmISBN{}
\acmDOI{}

\begin{document}

\author{Guilherme Moraes Rosa}
\affiliation{
  \institution{NeuralMind}
  \institution{University of Campinas (Unicamp)}
  \country{}
}

\author{Ruan Chaves Rodrigues}
\affiliation{
  \institution{NeuralMind}
  \institution{Federal University of Goiás (UFG)}
  \country{}
}

\author{Roberto de Alencar Lotufo}
\affiliation{%
  \institution{NeuralMind}
  \institution{University of Campinas (Unicamp)}
  \country{}
}

\author{Rodrigo Nogueira}
\affiliation{%
  \institution{NeuralMind}
  \institution{University of Campinas (Unicamp)}
  \institution{University of Waterloo}
  \country{}
}

\renewcommand{\shortauthors}{Rosa et al.}
\renewcommand{\shorttitle}{Yes, BM25 is a Strong Baseline for Legal Case Retrieval}

\begin{abstract}
We describe our single submission to task 1 of COLIEE 2021. Our vanilla BM25 got second place, well above the median of submissions.
Code is available at~\url{https://github.com/neuralmind-ai/coliee}.
\end{abstract}

\maketitle

\section{Introduction}

The Competition on Legal Information Extraction/Entailment (COLIEE)~\citep{rabelo2020coliee,rabelo2019coliee,kano2018coliee,Kano2017OverviewOC} is an annual competition to evaluate automatic systems on case and statute law tasks. 

In this paper, we describe our submission to the legal case retrieval task of COLIEE 2021.
The goal of this task is to explore and evaluate the performance of legal document retrieval technologies. It consists of retrieving from a corpus the cases that support or are relevant to the decision of a new case. These relevant cases are referred to as ``noticed cases''.

\section{Related Work}

Some successful NLP approaches to the legal domain use a combination of data-driven methods and hand-crafted rules~\cite{zhong2020does}. For example, in task 1 of COLIEE 2019, \citet{itp-2019} used a combination of techniques, such as Doc2Vec and BM25. \citet{ub-2020} used a learning to rank approach with features generated from models such as BM25 and TF-IDF. For task 1 of COLIEE 2020, \citet{iiest-2020} applied filtered-bag-of-ngrams and BM25. 

\citet{br-law} compared TF-IDF, BM25 and Word2Vec models for jurisprudence retrieval. The results indicated that the Word2Vec Skip-Gram model trained on a specialized legal corpus and BM25 yield similar performance. \citet{patent} investigate BERT \cite{devlin2019bert} for document retrieval in the patent domain and found that BERT model does not yet achieve performance improvements for patent document retrieval compared to the BM25 baseline.

\citet{pradeep5h2oloo} showed that BM25 is above the median of competition submissions in TREC 2020 Health Misinformation and Precision Medicine Tracks.

\section{The Task}

The dataset for task 1 is composed of predominantly Federal Court of Canada case laws, and it is provided as a pool of cases containing 4415 documents. The input is an unseen legal case, and the output is the relevant cases extracted from the pool that support the decision of the input case. The training set includes 650 query cases and 3311 relevant cases with an average of 5.094 labels per example. In the test set, only the query cases are given, 250 documents in total. We also show the statistics of this dataset in Table~\ref{table:task1}.

The micro F1-score is the official metric in this task:
\begin{equation}
    \text{F1} = (2 \times P \times R) / (P + R),
\end{equation}
\noindent where $P$ is the number of correctly retrieved cases for all queries divided by the number of retrieved cases for all queries, and $R$ is the number of correctly retrieved cases for all queries divided by the number of relevant cases for all queries.

\begin{table}[h!]

\centering
 \begin{tabular}{l | c | c } 
 \toprule
\textbf{} & \textbf{Train} &  \textbf{Test}  \\
 \midrule
 Number of base cases & 650 & 250  \\ 
 
 Number of candidate cases & 4415 & 4415 \\
 
 Number of relevant cases & 3311 & 900 \\
 
 Avg. relevant cases per base case & 5.1 & 3.6 \\
 
 \bottomrule
\end{tabular}
\vspace{0.1cm}
\caption{COLIEE 2021 task 1 data statistics.}
\label{table:task1}
\end{table}

\section{Our method: BM25}

BM25 \cite{rob1994,crestani1999} is an algorithm developed in the 1990s based on a probabilistic interpretation of how terms contribute to the relevance of a document and uses easily computed statistical properties such as functions of term frequencies, document frequencies and document lengths. The algorithm is a weighting scheme in the vector space model characterized as unsupervised, although it contains the free parameters $k_1$ and $b$ that can be tuned to improve results. 

BM25 score between a query $q$ and a document $d$ is derived from a sum of contributions from each query term that appears in the document and it is defined as:

\begin{multline}
    \textrm{BM25}(q, d) =\\ \sum_{t \in q \cap d} \log{\frac{N-\textrm{df}(t)+0.5}{\textrm{df}(t)+0.5}} \cdot \frac{\textrm{tf}(t,d) \cdot (k_1 + 1)}{\textrm{tf}(t,d) + k_1 \cdot \left( 1-b + b \cdot \frac{l_d}{L} \right)}
\end{multline}

The first part of the equation (the log term) is the inverse document frequency (idf):  $N$ is the total number of documents in the corpus, and $df(t)$ refers to the document frequency or the number of documents that term $t$ appears. In the second part, $tf(t, d)$ represents the number of times term $t$ appears in document $d$ or its term frequency. The denominator performs length normalization since collections usually have documents with different lengths. $ld$ refers to the length of document $d$ while $L$ is the average document length across all documents in the collection. As said before, $k_1$ and $b$ are free parameters.

Until today, BM25 still provides competitive performance in comparison with modern approaches on text ranking tasks.

We use BM25 from Pyserini, which is a Python library designed to help research in the field of information retrieval. It includes sparse and dense representations ~\cite{lin2021pyserini}. Pyserini was created to provide easy-to-use information retrieval systems that could be combined in a multi-stage ranking architecture in an efficient and reproducible manner. The library is self-contained as a standard Python package and comes with queries, pre-built indexes, relevance judgments, and evaluation scripts for many used IR test collections such as MS MARCO \cite{MS_MARCO_v3}, TREC \cite{pradeep5h2oloo, Roberts2019OverviewOT, zhang2020rapidly} and more. In this work, we use retrieval with sparse representations and it is provided via integration with Anserini \cite{anserini}, which is built on Lucene \cite{lucene}.

To apply BM25 to task 1, we first index all base and candidate cases present in the dataset. Before indexing, we segment each document into segments of texts using a context window of 10 sentences with overlapping strides of 5 sentences. We refer to these segments as candidate case segments.

In task 1, queries are base cases, which are also long documents. In our experiments, we found that using shorter queries improves efficiency and effectiveness. Thus, we apply to the base cases the same segmentation procedure described during the indexing step, creating, as we refer to, base case segments. We then use BM25 to retrieve candidate case segments for each base case segment.
We denote $s(b_i, c_j)$ as the BM25 score of the $i$-th segment of the base case $b$ and the $j$-th segment of the candidate case $c$.

The relevance score $s(b, c)$ for a (base case, candidate case) pair is the maximum score among all their base case segment and candidate case segment pairs:
\begin{equation}
    s(b, c) = \max_{i,j} s(b_i, c_j)
\end{equation}

We then rank the candidates of each base case according to these relevance scores and use the method described in Section~\ref{section:answer_selection} to select the candidate cases that will comprise our final answer.

Due to the large number of segments produced from base cases, retrieving the base cases of the test set takes more than 24 hours on a 4-core machine. Thus, we also evaluate our system using only the first $N$ segments. Table~\ref{table:task_1} summarizes our three best hyperparameters. The models are named using the format BM25-($N$, window size, stride). We achieve the best result using all base case segments, a window size of 10 sentences, and a stride of 5 sentences. However, due to the high computational cost of scoring all segments, our submitted system uses only the first 25 windows of each base case, i.e., $N=25$.

\begin{table}[h!]

\centering
 \begin{tabular}{l | c | c | c } 
 \toprule
\textbf{Method} & \textbf{F1} &  \textbf{Precision} &  \textbf{Recall} \\
 \midrule
 BM25-(10, 10, 5)  & 0.1040 & 0.0785 & 0.1560 \\
 BM25-(25, 10, 10) & 0.1203 & 0.0997 & 0.1516  \\ 
 BM25-(All, 10, 5)  & 0.1386 & 0.1027 & 0.2134 \\
 \bottomrule
\end{tabular}
\vspace{0.1cm}
\caption{\label{tab:task_1} Task 1 results on the 2021 dev set.}
\label{table:task_1}
\end{table}

\subsection{Answer Selection}
\label{section:answer_selection}

Given a base case $b$, BM25 estimates a relevance score $s(b, c)$ for each candidate case $c$ retrieved from the corpus using the method explained above.
To select the final set of candidate cases, we apply three rules:
\begin{itemize}
    \item Select candidates whose relevance scores are above a threshold $\alpha$;
    \item Select the top $\beta$ candidate cases with respect to their relevance scores;
    \item Select candidate cases whose scores are at least $\gamma$ of the highest relevance score.
\end{itemize}

\noindent We use an exhaustive grid search to find the best values for $\alpha$, $\beta$, $\gamma$ on the first 100 examples of the 2021 training dataset. 
We swept $\alpha = [0, 0.1, ..., 0.9]$, $\beta = [1, 5 ..., 200]$, and $\gamma = [0, 0.1, ..., 0.9, 0.95, 0.99,\\ 0.995, ..., 0.9999]$.

Note that our hyperparameter search includes the possibility of not using the first or third strategies if $\alpha=0$ or $\gamma=0$ are chosen, respectively.

\section{Results}  

\begin{table}[h!]

\centering
 \begin{tabular}{l | c | c | c | c } 
 \toprule
\textbf{Results} & \textbf{F1} &  \textbf{Precision} &  \textbf{Recall} \\
 \midrule
 Median of submissions & 0.0279 & - & -  \\
 3rd best submission of 2021 & 0.0456 & - & -  \\
 Best submission of 2021 & 0.1917 & - & -  \\
 \midrule 
 BM25 (ours) & 0.0937 & 0.0729 & 0.1311  \\ 
 \bottomrule
\end{tabular}
\vspace{0.1cm}
\caption{\label{tab:task_1_result} Task 1 results on the 2021 test set.}
\label{table:task_1_result}
\end{table}

Results are shown in Table \ref{tab:task_1_result}.
Our vanilla BM25 is a good baseline for the task as it achieves second place in the competition and its F1 score is well above the median of submissions. This result is not a surprise since it agrees with results from other competitions, such as the Health Misinformation and Precision Medicine tracks of TREC 2020~\cite{pradeep5h2oloo}. The advantage of our approach is the simplicity of our method, requiring only the document's segmentation and the grid search. One of the disadvantages is the time spent during the retrieval of segmented documents.

\section{Conclusion}

We showed that our simple BM25 approach is a strong baseline for the legal case retrieval task.

\subsubsection*{Acknowledgments}
This research was funded by a grant from Fundação de Amparo à Pesquisa do Estado de São Paulo (FAPESP) 2020/09753-5.


\bibliography{acmart}
\bibliographystyle{ACM-Reference-Format}


\end{document}